\documentclass[aps,prb,twocolumn,twoside,preprintnumbers,amsmath,amssymb,superscriptaddress,floatfix]{revtex4}
\usepackage{times}
\usepackage{graphicx}
\usepackage{amsfonts}
\usepackage{amsmath, amsthm, amssymb}
\usepackage{dsfont}
\usepackage{color}
\usepackage{soul} 

\newcommand{\e}[1]{\mathrm{e}^{#1}}

\newcommand{\eg}{\textit{e.g. }}

\begin{document}

\title{Giant Triplet Proximity Effect in $\pi$-biased Josephson Junctions with Spin-Orbit Coupling}

\author{Sol H. Jacobsen$^{\!1}$ and Jacob Linder}

\affiliation{Department of Physics, Norwegian University of
Science and Technology, N-7491 Trondheim, Norway}

\begin{abstract}
In diffusive Josephson junctions the phase-difference $\phi$ between the superconductors strongly influences the spectroscopic features of the layer separating them. The observation of a uniform minigap and its phase modulation were only recently experimentally reported, demonstrating agreement with theoretical predictions up to now - a vanishing minigap at $\phi=\pi$. Remarkably, we find that in the presence of intrinsic spin-orbit coupling a \textit{giant proximity effect due to spin-triplet Cooper pairs can develop at $\phi=\pi$}, in complete contrast to the suppressed proximity effect without spin-orbit coupling. We here report a combined numerical and analytical study of this effect, proving its presence solely based on symmetry arguments, which makes it independent of the specific parameters used in experiments. We show that the spectroscopic signature of the triplets is present throughout the entire ferromagnetic layer. Our finding offers a new way to artificially create, control and isolate spin-triplet superconductivity.

\end{abstract}

\date{\today}

\maketitle

\section{Introduction}
Spin-polarized superconductivity in which the Cooper pairs reside in a so-called triplet state is currently attracting much attention, topical examples including chiral $p$-wave pairing \cite{maeno}, singlet/triplet mixing in superconductors without an inversion center \cite{gorkov, bauer}, intrinsic coexistence of ferromagnetism and superconductivity in heavy-fermion compounds \cite{saxena, aoki}, and lately topological superconductivity \cite{fu, sau, kane_review, zhang_review}. In all of these cases, triplet Cooper pairs emerge which offer the tantalizing prospect of combining superconductivity and spintronics \cite{linder_natphys_14}.

Spinless singlet Cooper pairs can penetrate into other materials via tunneling from a superconductor across their mutual interface, a phenomenon know as the proximity effect \cite{buzdin_rmp_05}. When the adjacent material is a ferromagnet, the penetration distance depends on the texture of the magnetic exchange field, which may convert the singlet pairs into triplets. If triplet components with spin projection parallel to the field are generated, these correlations can penetrate much further into the ferromagnet, being known as the long-range triplet (LRT) component \cite{bergeret_rmp_05}. The primary route to creating the LRT component has so far been to use magnetically inhomogeneous structures\cite{bergeret_rmp_05, impurities} or non-equilibrium distribution functions and intrinsic triplet superconductors\cite{alternatives}. However, it was recently established that spin-orbit (SO) coupling can also provide the necessary rotation \cite{BergeretTokatly, konschelle} to create LRT. Experimentally, such SO coupling can be provided either by the structural asymmetry in thin-film hybrid structures \cite{Edelstein2003}, giving rise to generic interfacial spin-currents \cite{LinderYokoyama2011}, or due to the intrinsic crystal structure, i.e. provided by noncentrosymmetric materials \cite{Samokhin2009}. 

The superconducting proximity effect is a phase-coherent phenomenon that can be probed in \eg Josephson junctions where the density of states is highly sensitive\cite{hammer_prb_07} to the superconducting phase difference $\phi$. Experimentally, Ref. \onlinecite{leSueuretal2008} reported measurements for Josephson junctions with a normal metal (SNS), that were consistent with the theoretical prediction \cite{zhou_prb_98} that the density of states evolves from a finite minigap due to the superconducting correlations $(\phi=0$) to the absence of such a minigap ($\phi=\pi$). This can be understood intuitively: the proximity effect would be suppressed when the order parameter in each superconductor is equal in magnitude but opposite in sign, resulting  in superconducting correlations ``averaging" to zero and a featureless density of states in the center of the normal-metal region.

\begin{figure}[htb]
  \centering
  \includegraphics[width=7cm]{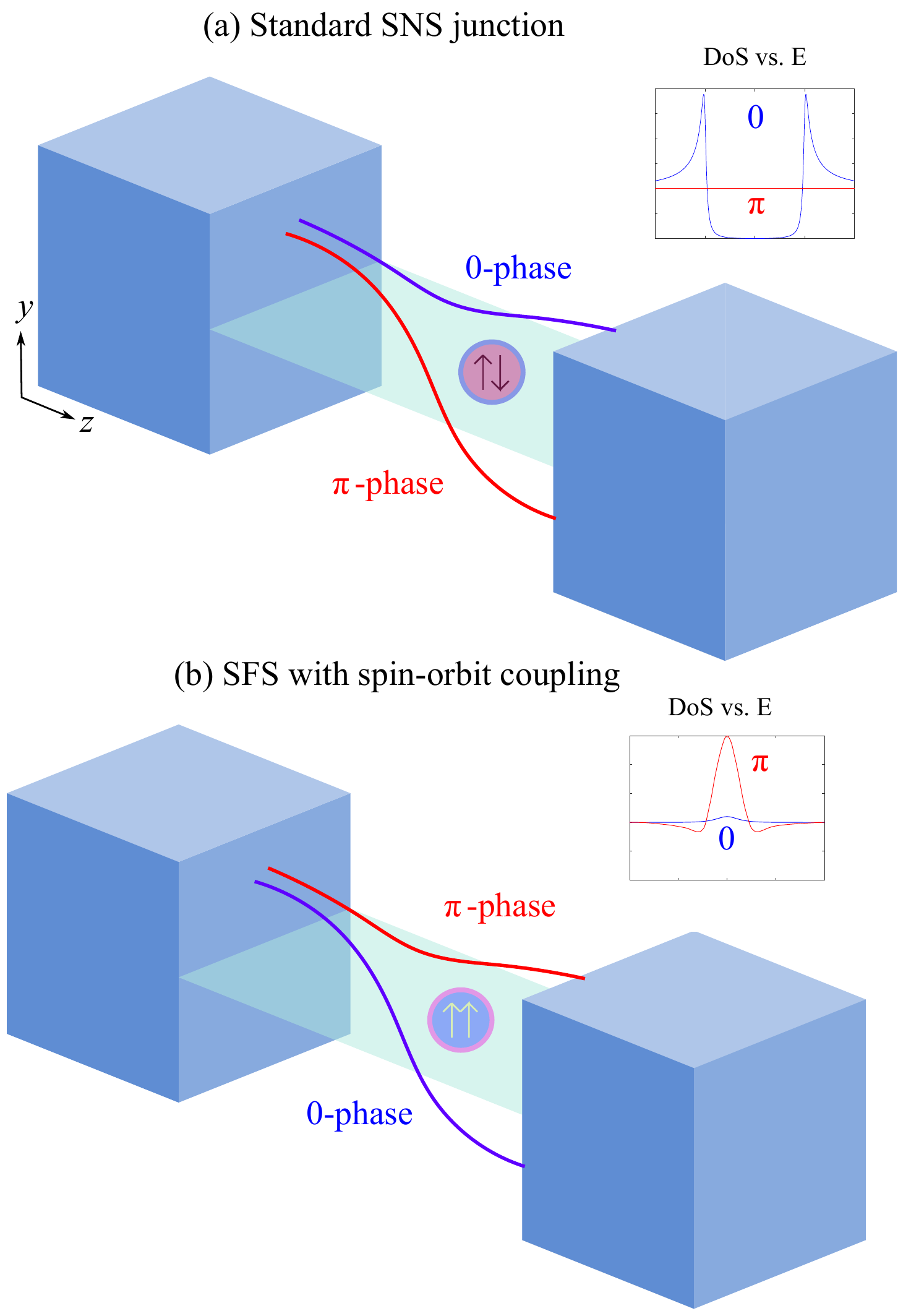}
  \caption[SFS Junction]
   {The proximity effect in the standard SNS Josephson junction (a) manifests in a minigap that closes with increasing phase difference $\phi$ between the superconductors, closing entirely for $\phi=\pi$. Conversely, the proximity effect in the SFS junction with intrinsic SO coupling in the junction direction (b) results in a giant peak in the density of states at zero energy for $\phi=\pi$. The different behaviour is due to the symmetry of the anomalous Green's function: for SNS, the singlet component is symmetric around the centre of the junction at $\phi=0$ and antisymmetric for $\phi=\pi$ whereas the SFS case retains a symmetric triplet component at $\phi=\pi$. For the examples used in-text we take the junction direction to lie in the $z$-direction and use bulk values for the superconductors. }
	\label{Fig:model}
\end{figure}

We here discover that when SO coupling is present in a magnetic Josephson junction, the opposite effect takes place. Remarkably, the SO coupling in the junction instead gives rise to a giant, triplet-induced proximity effect at $\phi=\pi$, as shown schematically in Fig. \ref{Fig:model}. We prove that the reason for this is that the SO coupling forces the triplet Cooper pairs to have the opposite parity symmetry compared to the singlet Cooper pairs with respect to the center of the junction. When $\phi=\pi$, the singlet correlations are antisymmetric across the junction whereas the triplets are symmetric, meaning that the proximity effect survives even in the centre of the junction and is solely due to LRT Cooper pairs. In other words the phase difference, which is an experimentally tunable quantity, can be used to remove the spin-singlets and keep only triplets even with a homogeneous exchange field. Previous attempts to separate spin-polarized Cooper pairs from the singlet component have required magnetic inhomogeneities, which are experimentally challenging to control, so the inclusion of SO coupling for this purpose represents a significant step forwards. We prove analytically, solely from symmetry arguments, that the giant triplet proximity effect always occurs near the centre of the junction, making its existence independent of the specific junction parameters used. Whereas the density of states can be probed locally, as can spin injection and tunneling in superconductors \cite{Fischer2007}, we also show numerically that the generation of spin-polarized Cooper pairs for $\pi$-biased Josephson junctions in fact persists throughout the entire system. The fact that our prediction is based solely on symmetry and that its spectroscopic fingerprint is independent of where the local density of states is measured in the system makes it a very robust effect.

The remainder of the article will be organised as follows: In Section \ref{Sec:Theory} we introduce the quasiclassical theory and notation for the trilayer Josephson junctions and discuss the analytic consequences of the inclusion of SO coupling in this formalism. In Section \ref{Sec:Results} we present numerical results demonstrating the emergence of the giant triplet proximity effect at $\phi=\pi$, and moreover show that this persists throughout the interstitial junction layer. Finally, we conclude in Section \ref{Sec:Disc} with a brief summary and discussion.

\section{Theory}\label{Sec:Theory}
It is instructive to briefly consider the behavior of the proximity effect in an SFS junction without SO coupling as a function of the phase difference. In this case the quasiclassical Usadel equation \cite{Usadel1970} in the linearized regime without SO coupling reads $D_F \partial_z^2 f_\pm + 2\textit{i}\varepsilon_\pm f_\pm = 0$, where $\varepsilon_\pm = \epsilon\pm h_z$ and $f_\pm = f_t \pm f_s$ for energy $\epsilon$, magnetisation exchange field $h$, diffusion constant $D_F$ in the ferromagnet and singlet and triplet anomalous Green's functions $f_s$ and $f_t$ respectively. The Usadel equation describes the diffusion of the condensate into the adjacent material, and the corresponding Kupriyanov-Lukichev boundary conditions \cite{KuprianovLukichev1988} to the superconducting interfaces take the form $\zeta L_F\partial_z f_\pm = \mp f_\text{BCS}\e{\textit{i}\phi_L}$ at $z=0$ and $\zeta L_F\partial_z f_\pm = \pm f_\text{BCS}\e{\textit{i}\phi_R}$ at $z=L_F$ where $f_\text{BCS}$ is the bulk, Bardeen-Cooper-Schrieffer anomalous Green's function in the superconductors, $L_F$ is the length of the ferromagnet and $\zeta$ is the interface parameter. $\phi_L$ and $\phi_R$ denote the respective superconducting phases. The solution in the middle of the junction reads:
\begin{align}
f_\pm = \frac{\pm f_\text{BCS} \cos(k_\pm L/2)}{\zeta L \textit{i} \sin(k_\pm L)}(\e{\textit{i}\phi_R} + \e{\textit{i}\phi_L}),
\end{align}
where $k_\pm=\sqrt{2 i\varepsilon_\pm/D_F}$ is the wavenumber. By direct insertion, one observes that when the phase difference $\phi=\phi_L-\phi_R=\pi$, the superconducting proximity effect vanishes completely since $f_\pm=0$. This holds for all energies and regardless of whether $h=0$ or $h\neq 0$.  Since this takes place at $\phi=\pi$ both in SNS and SFS junctions, one might be led to think that this is a robust phenomenon. However, we now show that in the presence of spin-orbit interactions, this is no longer the case. Not only is there a strong superconducting proximity effect present in the junction at $\phi=\pi$, but for certain cases it is in fact the \textit{maximum triplet proximity effect} that can be obtained in the junction. This is in stark contrast to the conventional picture of a vanishing minigap at $\phi=\pi$ reported previously for SNS junctions \cite{leSueuretal2008}.

The giant triplet proximity effect can be established solely on symmetry arguments, making it independent of the specific junction parameters employed in an experiment. To see this, we have derived the Usadel equations in terms of a Ricatti-parametrization \cite{schopohl} making it very suitable for numerical calculations, which we use later in this paper. The Usadel equation with SO coupling expressed in terms of the full Green's function was derived in Ref.~\onlinecite{BergeretTokatly} -- in its Ricatti-parametrized form in the ferromagnet, one obtains (see Ref.~\onlinecite{JacobsenOuassouLinder2015} for our full derivation):
\begin{eqnarray}\label{Eqn:SOUsadel}
\lefteqn{D_F\left(\partial_k^2 \gamma + 2(\partial_k \gamma)\tilde{N}\tilde{\gamma}(\partial_k \gamma)\right)} \nonumber\\
&=&\!\! -2i\epsilon\gamma - i\underline{h}.(\underline{\sigma}\gamma-\gamma\underline{\sigma}^*)\nonumber\\
&&\!\!+D_F\!\!\left[\!\underline{A}\underline{A}\gamma\!-\!\gamma\underline{A}^*\underline{A}^*\!+\!2(\underline{A}\gamma\!+\!\gamma\underline{A}^*)\tilde{N}(\underline{A}^*\!+\!\tilde{\gamma}\underline{A}\gamma)\!\right]\nonumber\\
&&\!\!+2iD_F\!\!\left[\!(\partial_k \gamma)\tilde{N}(\hat{A}^*_k\!+\!\tilde{\gamma}\hat{A}_k\gamma)\!+\!(\hat{A}_k\!+\!\gamma \hat{A}^*_k\tilde{\gamma})N(\partial_k \gamma)\!\right]\!\!.\,\,\,\,\,\,
\end{eqnarray}
Here $\underline{A}=(\hat{A}_x,\hat{A}_y,\hat{A}_z)$ is the SO field with components $\hat{A}_k=A_k^a\sigma^a$, using the summation convention over repeated indices with $a=x,y,z$, and index $k$ indicates an arbitrary choice of direction in Cartesian coordinates. The vectors $\underline{h}$ and $\underline{\sigma}$ are the corresponding three-component exchange field and vector of Pauli matrices. The $\gamma$, $\tilde{\gamma}$ are the Ricatti-parametrised matrices characterising the quasiclassical Green's function $\hat{g}^R$:
\begin{eqnarray}\label{Eqn:gR}
\hat{g}^R=
\begin{pmatrix}
N(\textit{I}+\gamma\tilde{\gamma}) & 2N\gamma\\
-2\tilde{N}\tilde{\gamma} & -\tilde{N}(\textit{I}+\tilde{\gamma}\gamma)
\end{pmatrix},
\end{eqnarray}
with normalisation matrices $N=(\textit{I}-\gamma\tilde{\gamma})^{-1}$ and $\tilde{N}=(\textit{I}-\tilde{\gamma}\gamma)^{-1}$ and identity matrix $I$. The $\tilde{\cdot}$ operation denotes complex conjugation and $\epsilon\rightarrow(-\epsilon)$. Similarly, the boundary conditions become:
\begin{eqnarray}
\partial_k\gamma_1=\frac{1}{L_1\zeta_1}(\textit{I}-\gamma_1\tilde{\gamma}_2)N_2(\gamma_2-\gamma_1)+i\hat{A}_k\gamma_1+i\gamma_1\hat{A}_k^*,\nonumber\\
\partial_k\gamma_2=\frac{1}{L_2\zeta_2}(\textit{I}-\gamma_2\tilde{\gamma}_1)N_1(\gamma_2-\gamma_1)+i\hat{A}_k\gamma_2+i\gamma_2\hat{A}_k^*.
\end{eqnarray}

In the last line of Eq. (\ref{Eqn:SOUsadel}) we see that the first order derivative couples to the component of the SO field in the corresponding direction. This is the key requirement for the appearance of the giant triplet proximity effect at $\phi=\pi$. Consider therefore an SFS junction oriented along the $z$-direction as in Figure \ref{Fig:model}, and a choice of SO coupling vector aligned perpendicular to the interfaces ($\hat{A}_z\neq0$). For concreteness and to give more transparent analytical results, we set $\underline{A}=(0,0,\alpha\sigma_x-\alpha\sigma_y)$. This choice corresponds \eg to Rashba-type coupling with broken inversion symmetry in the $\hat{n}=\hat{x}+\hat{y}$ direction, with the transverse motion of the electrons restricted, effectively corresponding to motion in a one-dimensional wire. We underline that the non-vanishing triplet proximity effect in $\pi$-biased Josephson junctions predicted here survives even if one were to include components of the SO coupling field $\{\hat{A_x},\hat{A_y}\}\neq 0$ and if one were to only include the $\sigma_x$ or $\sigma_y$ term in $\hat{A_z}$ above.  We can then write down the linearised Usadel equations in the limit of weak proximity effect, where we have $|\gamma|_{ij}\ll 1$, $N\approx 1$ in the ferromagnet such that 
\begin{eqnarray}\label{Eqn:wpegamma}
2\gamma=
\begin{pmatrix}
f_{\uparrow\uparrow} & f_{s}+f_{t}\\
f_{t}-f_{s} & f_{\downarrow\downarrow}
\end{pmatrix}.
\end{eqnarray}
The origin and main features of the giant triplet proximity effect can now be identified analytically by considering the low-energy regime $\varepsilon=0$ and setting the exchange field to $\underline{h}=h_z\hat{z}$, where the equations become:
\begin{align}
\label{Eqn:AzWPE}
&(\partial^2_z-4\alpha^2) f_{\sigma\sigma} + 4\sigma\alpha(1-\sigma i)\partial_z f_t -4i\sigma \alpha^2 f_{-\sigma,-\sigma} =0,\notag\\
&D_F\partial^2_z f_s+2ih_zf_t =0,\notag\\ 
&D_F\partial^2_z f_t + 2ih_zf_s - 8D_F\alpha^2 f_t \notag\\
&+ 2D_F\alpha(1-i)\partial_z f_{\downarrow\downarrow} - 2D_F\alpha(1+i)\partial_z f_{\uparrow\uparrow} = 0.
\end{align}
with $\sigma=\uparrow,\downarrow$. When there is zero phase difference between the superconductors, the anomalous Green's function for the singlet pairs, $f_s$, is a symmetric function with respect to the middle of the junction. This can be seen immediately from the general form of the solution of $f_s$ and the boundary conditions, and is equivalent to what happens for conventional SNS and SFS junctions. When the phase difference is equal to $\pi$, however, $f_s$ (and thus necessarily its second derivative) is antisymmetric. Now, it follows from Eq. \ref{Eqn:SOUsadel} that when the SO coupling has a component in the junction direction it necessarily introduces a first-order derivative term. Performing the operation $z\to (-z)$ on Eq. (\ref{Eqn:AzWPE}), this means that the function subject to the first order differential must be a symmetric function, provided it is not constant. In the linearised regime (\ref{Eqn:AzWPE}) we can see explicitly what this entails: the functions $f_{\uparrow\uparrow}$ and $f_{\downarrow\downarrow}$ must be symmetric around the middle of the junction, and it is clear that a nonzero component of the anomalous Green's function will remain at zero energy even in the $\pi$-biased junction. Since the density of states in the ferromagnet is given by
\begin{eqnarray}\label{Eqn:DOS}
D(\!\epsilon) \!\!&=&\!\! \rm{Tr}\left\{\textit{N}(\textit{I}+\gamma\tilde{\gamma})\right\}/2,
\end{eqnarray}
such that the expression at zero energy in the linearised regime becomes
\begin{eqnarray}\label{Eqn:DOSzero}
D(0) \!\!&=&\!\!1\!-\!\frac{|f_s(0)|^2}{2} \!+\! \frac{|f_t(0)|^2}{2} \!+\! \frac{|f_{\uparrow\uparrow}(0)|^2}{4} \!+\! \frac{|f_{\downarrow\downarrow}(0)|^2}{4},\,\,\,
\end{eqnarray}
an experimental signature of this effect would be an enhanced zero-energy density of states due to the triplets. Without SO coupling, this typically occurs for exchange fields fulfilling the resonant condition\cite{Yokoyama2005,Kawabata2013} $h\sim E_g$, where $E_g$ is the minigap energy. However, the mechanism for the enhanced density of states due to triplets in the present work is fundamentally different from previous literature as it originates from a parity effect due to SO coupling which is present independently on the exact junction parameters. We underline that the purpose of Eq. (\ref{Eqn:DOSzero}) is to illustrate that the triplets will enhance the density of states \cite{buzdin_prb_00} whereas in the actual numerical calculations shown in the figures we have solved the non-linearized Usadel equation and computed the density of states via Eqn. (\ref{Eqn:DOS}). Although the precise phase difference that produces maximal peak at zero energy depends on the strengths of the exchange field and SO coupling (and we leave a more detailed exposition of the model dependencies for later work) the $\phi=\pi$ case remains significant and can also be maximal. We emphasize again that this discovery is the complete opposite of what has been thought to be the case up to now in Josephson junctions, namely a vanishing superconducting proximity effect whenever the phase difference is $\pi$.

\section{Results}\label{Sec:Results}
\begin{figure}[htb]
  \centering
  \includegraphics[width=\linewidth]{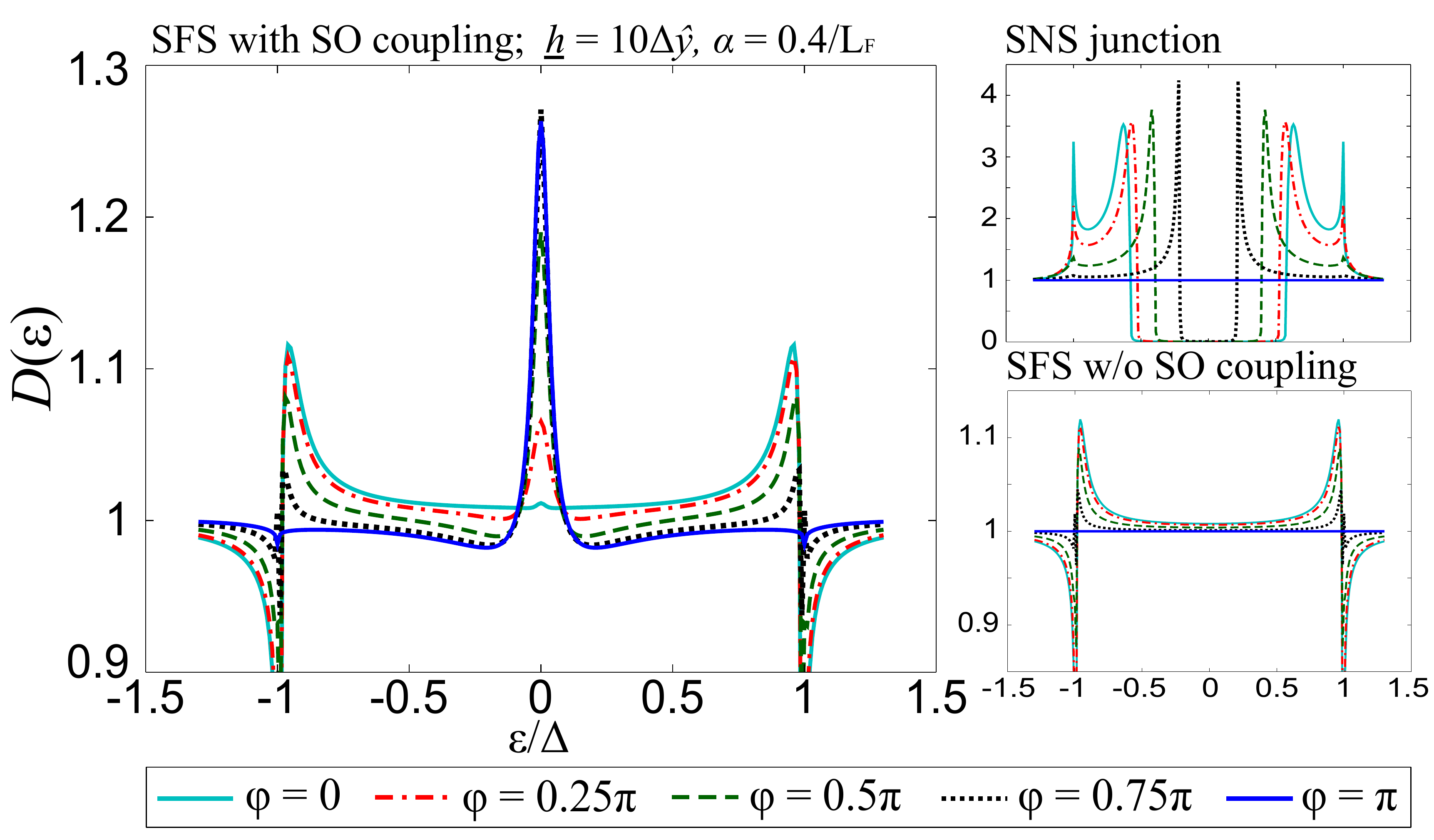}
  \caption
   {The local density of states $D(\epsilon)$ at $z=L_F/2$ for phase difference $\phi$ between the superconductors of an SFS junction with spin-orbit coupling aligned in the junction direction, with exchange field $\underline{h}=10\Delta\hat{y}$ and spin-orbit coupling $\alpha=0.4/L_F$. The giant triplet proximity effect at $\phi=\pi$ manifests as a large peak in the $D(\epsilon)$ at $\epsilon=0$. A comparison with the standard SNS and SFS junctions without SO coupling is also provided.}
	\label{Fig:DoSvE}
\end{figure}

In what follows, we present for the first time a solution to the Usadel equation with spin-orbit coupling in the full proximity regime. In Fig. \ref{Fig:DoSvE} we show an example of the spectroscopic profile for varying phase difference between the superconductors in an SFS junction with SO coupling, highlighting the generation of a zero-energy peak in the density of states at $\phi=\pi$. We choose an in-plane exchange field $\underline{h}=10\Delta \hat{y}$ for ease of experimental application, a bulk superconducting coherence length $\xi_S = 30$ nm and SO coupling strength $\alpha=0.4/L_F$, i.e. normalised to the ferromagnet length $L_F$, here chosen to be $15$ nm such that the relevant quantity $L_F/\xi_S=0.5$. Qualitatively similar behaviour is observed for most other choices of exchange field orientation and SO coupling strength. A comparison with the standard SNS and SFS junctions without SO coupling is provided and the giant proximity effect at $\phi=\pi$ is immediately clear. Since the singlet component $f_s$ vanishes when $\phi=\pi$, the remaining features are entirely due to the triplets and in this case entirely due to the LRT component. In effect, the result reported here serves as a way to fully isolate the triplet correlations regardless of the junction parameters in $\pi$-biased Josephson junctions. We note that quantitatively, even when $h\gg\Delta$, the proximity effect and resulting enhancement of the density of states here is much larger than experiments measuring the same quantity for superconductor-ferromagnet hybrids without SO coupling \cite{kontos_prl_01, sangiorgio_prl_07}, where the deviation from the normal-state is about 1\%.

\begin{figure}[t!]
\centering
\resizebox{0.45\textwidth}{!}{
\includegraphics{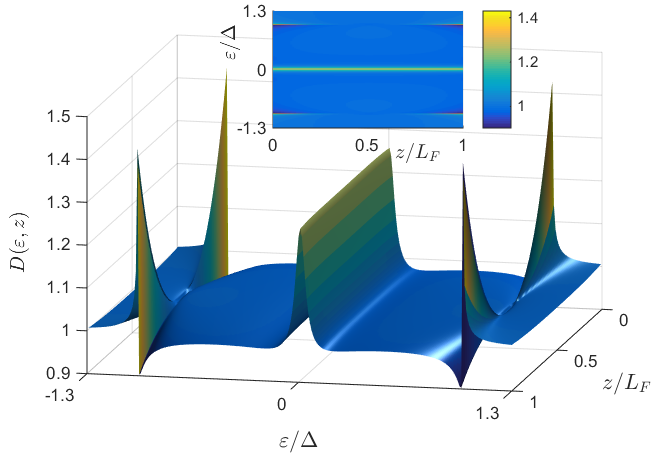}}
\caption{(Color online) The spatial dependence of the local density of states $D(\epsilon,z)$ for a phase differences $\phi=\pi$ between the superconductors of an SFS junction with spin-orbit coupling. Parameter values are the same as in Fig. \ref{Fig:DoSvE}. As seen, the giant proximity effect is virtually independent on the position from the interfaces, thus persisting throughout the entire system. \textit{Inset:} top view of the DoS surface plot.}
\label{fig:spatial} 
\end{figure}

\begin{figure}[htb]
  \centering
  \includegraphics[width=0.9\linewidth,bb=0 0 1200 655,clip]{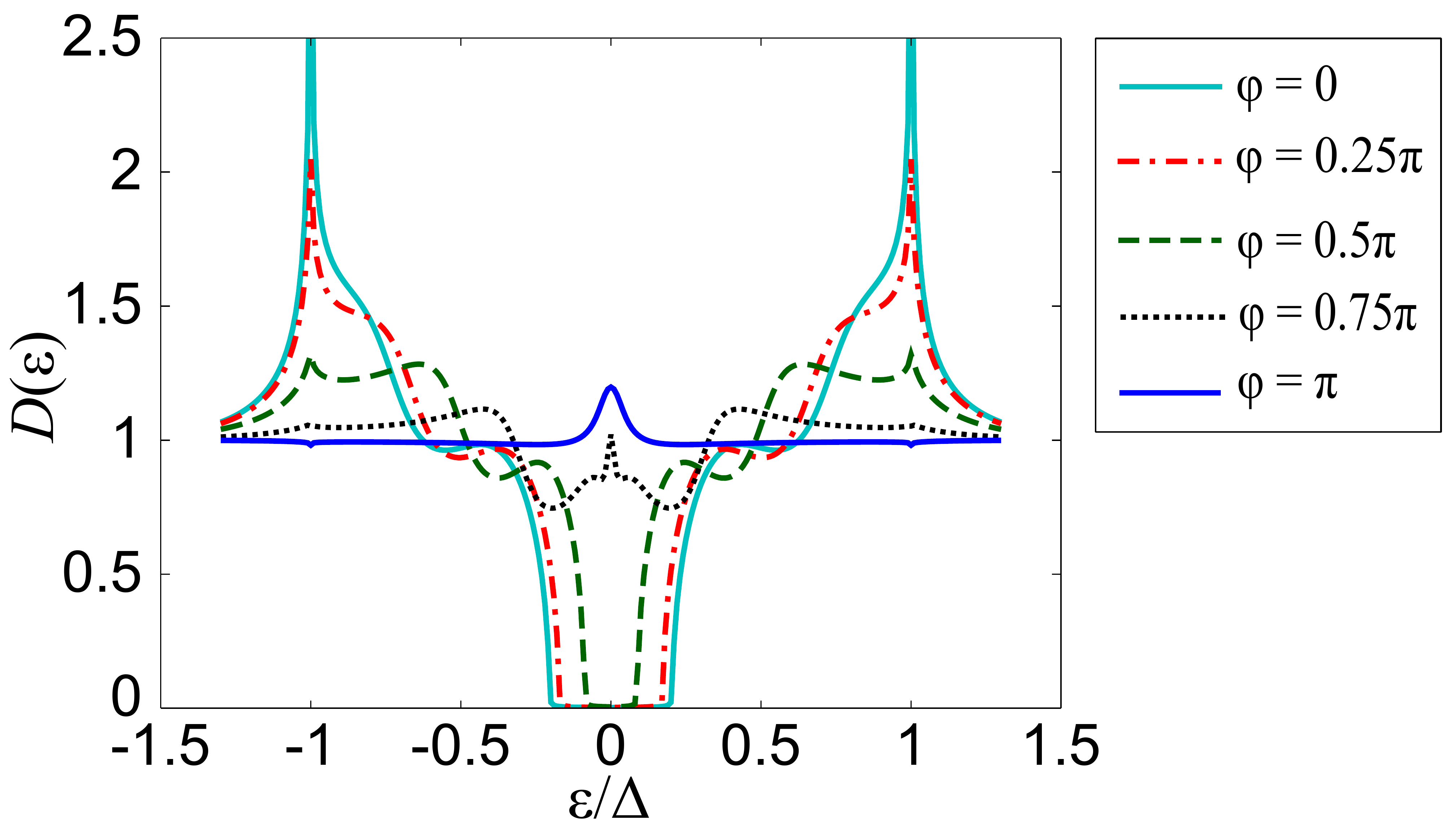}
  \caption
   {The local density of states $D(\epsilon)$  at $z=L_F/2$ for phase differences $\phi$ between the superconductors of an SFS junction with spin-orbit coupling aligned in the junction direction, with exchange field $\underline{h}=3\Delta\hat{z}$ and spin-orbit coupling $\alpha=2/L_F$. At small phase differences $D(\epsilon)$ displays a minigap while retaining a zero-energy peak at $\phi=\pi$.}
	\label{Fig:DOS_hz3_RD2}
\end{figure}

Another important question pertains to how the manifestation of the giant triplet proximity effect depends on the distance from the superconductors and also on asymmetries between the two interfaces due to \eg different transparencies. Remarkably, the effect is virtually independent of the distance from the superconducting interfaces: the spectroscopic peak originating from the presence of spin-polarized Cooper pairs persists all the way up to the interfaces and hardly changes throughout the junction as shown in Fig.~\ref{fig:spatial}. For reference, the spatial distribution of the anomalous Green's function is provided in Fig.~\ref{Fig:SpatialComps} in the Appendix. For substantially stronger spin-orbit coupling, a spatially oscillatory contribution from the short range triplet component with spin projection perpendicular to the field can also arise as one moves toward the interfaces, but the proximity effect still remains completely dominated by triplets throughout the entire system, i.e. not only near the middle of the junction (see Figs.~\ref{Fig:DOS_hz3_RD2} and \ref{Fig:SpatialComps}). Moreover, we have also checked numerically (not shown) that the spatial dependence remains unchanged even for asymmetric junctions where one interface is as much as twice as transparent as the other, i.e. a ratio of barrier parameters $\zeta_1/\zeta_2=2$. This suggests that the predicted effect should be very robust and facilitates its experimental observation.

\begin{figure}[htb]
  \centering
  \includegraphics[width=\linewidth]{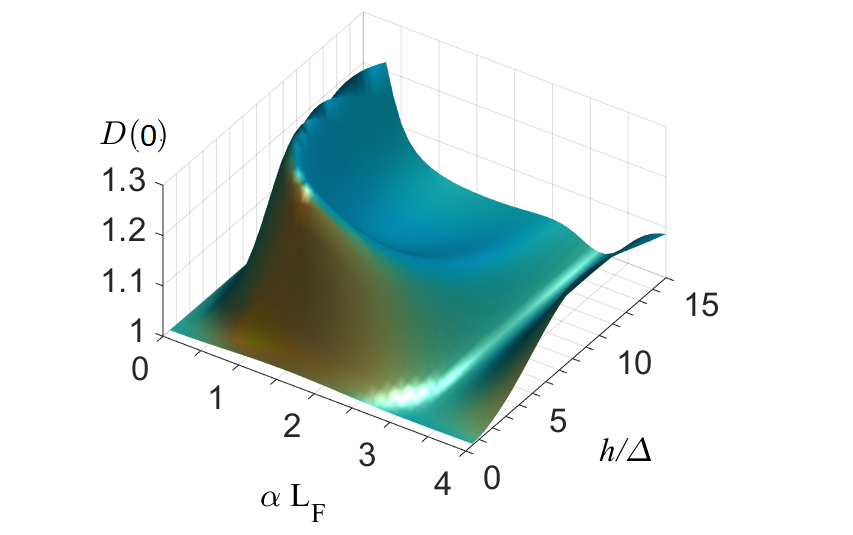}
  \caption
   {The local density of states $D(\varepsilon=0)$  at $z=L_F/2$ for a $\pi$-biased SFS junction with spin-orbit coupling vector aligned in the junction direction (corresponding to a nanowire setup) as a function of magnetization exchange field $\underline{h}=h\Delta\hat{y}$ and strength of spin-orbit coupling $\alpha L_F$ aligned in the junction direction.}
	\label{Fig:DoS_h_SOC}
\end{figure}

In Figure \ref{Fig:DoS_h_SOC} we show the intricate dependence of the density of states of the $\pi$-biased junction on the strengths of exchange field $\underline{h}=h\Delta\hat{y}$ and SO coupling $\alpha L_F$, which is highly nonmonotonic in $\alpha$. We see that as the field strength increases a more narrow spectrum of SO coupling will generate a giant peak at zero energy, with the optimal SO coupling decreasing slightly for higher field strengths. 

In Figure \ref{Fig:DoSvE} we saw that the peak at zero energy was present for all phases, increasing significantly as $\phi\rightarrow \pi$. This is the predominant behaviour observed for the majority of the $\alpha$-$\underline{h}$ parameter space, with a few exceptions. In particular, we note that when the exchange field is oriented in the junction direction, i.e. in the same direction as the SO coupling, increasing the SO coupling can take the density of states profile from having a zero-energy peak for all phase differences as in Fig.~\ref{Fig:DoSvE}, to a profile with a qualitative change from minigap at $\phi=0$ to peak at $\phi=\pi$, as shown in Fig.~\ref{Fig:DOS_hz3_RD2}. The reason for the appearance of a full minigap in Fig.~\ref{Fig:DOS_hz3_RD2} at zero phase difference in spite of the presence of an exchange field can be understood by noting that for large spin-orbit coupling, the triplet component becomes suppressed whereas the singlet one remains unaffected. In this case the SO coupling stabilizes the minigap, which now persists throughout the interstitial junction layer, and leads to a magnetically tunable minigap \cite{JacobsenOuassouLinder2015}. However, the isolation of spin-polarized LRT Cooper pairs at $\phi=\pi$ survives, providing an enhanced density of states $D(\varepsilon=0)>1$, indicating that it is a robust phenomenon.

\section{Discussion and Summary}\label{Sec:Disc}
In the present analysis we restricted the form of the SO coupling vector to lie along the junction direction, as in the case of one-dimensional wires, although we have also verified that the giant proximity effect is maintained for weak transverse SO coupling in the $xy$-plane. A common alternative way of introducing intrinsic SO coupling compared with using a noncentrosymmetric crystal is to use the interfacial asymmetry of thin-film junctions, and in this case Rashba-Dresselhaus coupling will lie in the $xy$-plane. This case will be explored in more detail in Ref.~\onlinecite{JacobsenOuassouLinder2015}. We also considered here the coexistence of the exchange field and the SO coupling, which may also be taken as a model of the case where these features exist in separate, adjoining layers, such that the SO coupling is induced, \eg by deposition of a heavy transition metal or compound. Such a setup could be easier to integrate into current devices, and the consequences for supercurrent generation will be explored in the future\cite{JKL2015}.

In summary, we have shown that the inclusion of intrinsic spin-orbit coupling in the ferromagnet of an SFS junction can result in a pure spin-triplet state without any singlet superconductivity, and that this state can persist throughout the ferromagnet. Remarkably, this occurs for a phase difference of $\pi$ between the superconductors, dispelling the commonly held view that the proximity effect should be suppressed at $\phi=\pi$. In fact, in many cases the peak at zero energy will be significantly enhanced compared with the features at zero phase difference, displaying what we term the giant proximity effect in $\pi$-biased junctions. The inclusion of intrinsic spin-orbit coupling is therefore a very promising resource for harnessing triplet superconductivity for improved functionality in spintronic devices.

\textit{Acknowledgments.} We thank J. A. Ouassou for very useful discussions. The authors acknowledge support from the Outstanding Academic Fellows programme at NTNU, the COST Action MP-1201 'Novel Functionalities through Optimized Confinement of Condensate and Fields', and the Norwegian Research Council Grant No. 205591 (FRINAT) and Grant No. 216700.

\appendix
\section{Spatial variation of anomalous Green's function}\label{App:comps}
\begin{figure*}[htb]
  \centering
  \includegraphics[width=\textwidth]{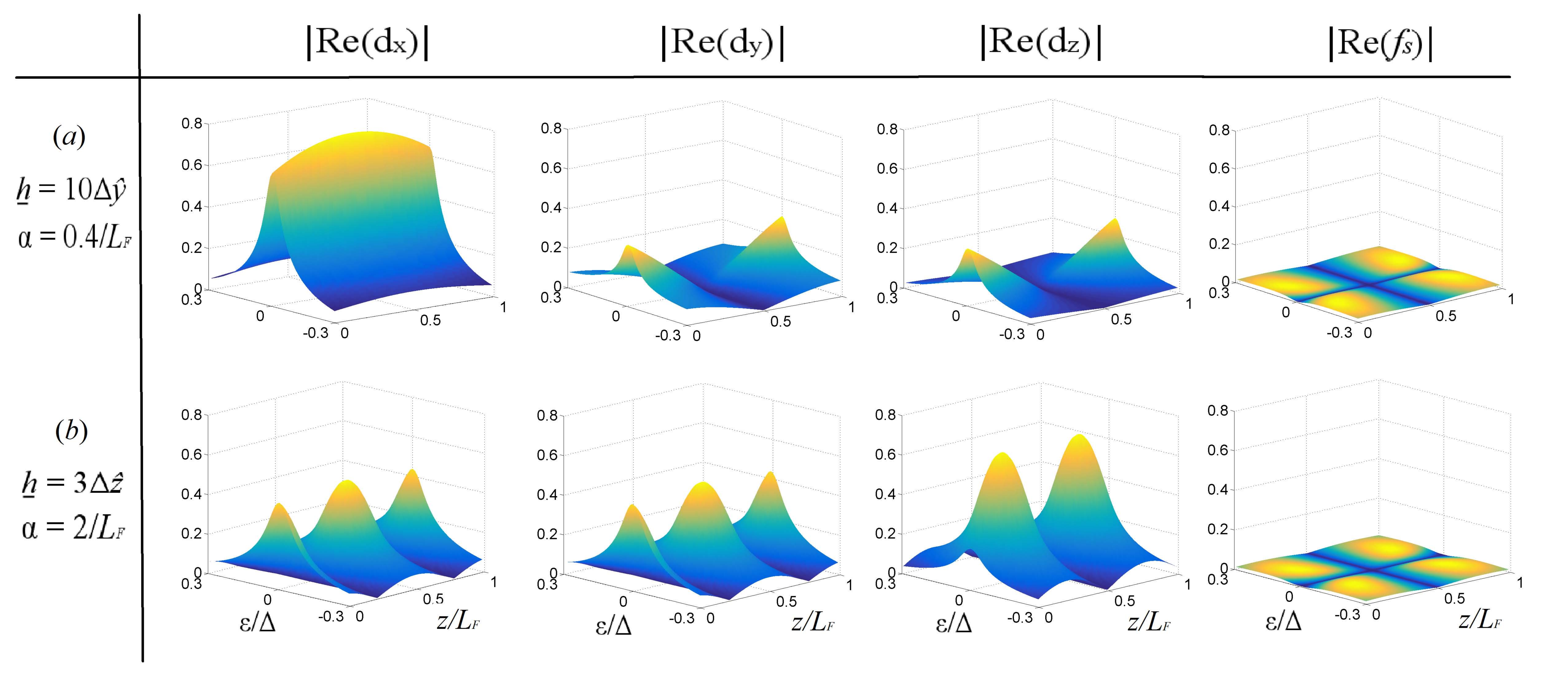}
  \caption
   {Spatial variation of the singlet and triplet components throughout the ferromagnet of the SFS junctions considered in the main text.}
	\label{Fig:SpatialComps}
\end{figure*}

To examine the spatial variation of the singlet and triplet components of the anomalous Green's function $f$, we write
\begin{equation}
	f = (f_s + \underline{d} \cdot \underline{\sigma}) i\sigma_y \, .
\end{equation}
Now the singlet component $f_s$ transforms as a scalar under spin rotations while the triplet component transforms as a vector\cite{BalianWerthamer1963, MackenzieMaeno2003} and is described by the $d$-vector $\underline{d}=(d_x,d_y,d_z)=((f_{\downarrow\downarrow}-f_{\uparrow\uparrow}), -i(f_{\downarrow\downarrow}+f_{\uparrow\uparrow}), 2f_t)/2$. In Fig. \ref{Fig:SpatialComps} we show how the magnitude of the real part of the singlet and triplet components vary throughout the ferromagnet for the examples discussed in the main text (the imaginary component is zero at $\varepsilon=0$). With an in-plane exchange field $\underline{h}=10\Delta\hat{y}$ as shown in Fig. \ref{Fig:SpatialComps}$(a)$, the LRT component is given by the dominant $|\textrm{Re}(d_x)|$ throughout the entire length of the ferromagnet. The short-ranged triplet component is defined as the component of the $d$-vector parallel to the exchange field, and it is evident that in this case the contribution from the short-ranged triplet increases towards the superconductor interfaces but remains minimal compared with the LRT component.

Increasing the strength of SO coupling can give rise to oscillations between the long- and short-ranged components, as is evident in Fig. \ref{Fig:SpatialComps}$(b)$, with an out-of-plane exchange field $\underline{h}=3\Delta\hat{z}$. In this case both $d_x$ and $d_y$ are symmetric and show the long-ranged component with a maximal peak around the middle of the junction. The short-ranged component $d_z$ is zero in the centre of the junction and increases correspondingly as the LRT decays. In both cases the singlet contribution is negligible, such that the triplet component dominates throughout the junction. The strength of the exchange field has a minimal effect on the frequency of oscillation in comparison with the strength of SO coupling.

\end{document}